\documentclass[12pt]{article}

\begin{document}

\tolerance=5000
\def\pp{{\, \mid \hskip -1.5mm =}}
\def\cL{{\cal L}}
\def\be{\begin{equation}}
\def\ee{\end{equation}}
\def\bea{\begin{eqnarray}}
\def\eea{\end{eqnarray}}
\def\beaa{\begin{eqnarray*}}
\def\eeaa{\end{eqnarray*}}
\def\tr{{\rm tr}\, }
\def\nn{\nonumber \\}
\def\e{{\rm e}}
\def\D{{D \hskip -3mm /\,}}

\def\SEH{S_{\rm EH}}
\def\SGH{S_{\rm GH}}
\def\AdS5{{{\rm AdS}_5}}
\def\S4{{{\rm S}_4}}
\def\gfv{{g_{(5)}}}
\def\gfr{{g_{(4)}}}
\def\SC{{S_{\rm C}}}
\def\RH{{R_{\rm H}}}

\def\wlBox{\mbox{
\raisebox{0.1cm}{$\widetilde{\mbox{\raisebox{-0.1cm}\fbox{\ }}}$}}}
\def\BBox{\mbox{\raisebox{0.1cm}\fbox{\ }}}
\def\htBox{\mbox{
\raisebox{0.1cm}{$\hat{\mbox{\raisebox{-0.1cm}{$\Box$}}}$}}}

\ 

\vfill

\begin{center}

{\bf\large Brane New World 
and dS/CFT correspondence}

\vfill

{\sl Sergei D. Odintsov}\footnote{Plenary talk at GRG11, Tomsk, July of 2002, 
email: odintsov@mail.tomsknet.ru} \\
{\it Labor. for Fundam. Study, Tomsk State Pedagogical University,
634041 Tomsk, RUSSIA\\}

\vfill

{\bf abstract}

\end{center}

 The occurence of 5d de Sitter space with 4d de Sitter brane 
is discussed on classical and quantum level. It is shown
that quantum effects maybe produced by dual CFT living on the brane.
Moreover, gravity trapping on the brane is proved via the 
presentation of 5d dS gravity as 4d gravity coupled with  gauge theory.
This supports the dS/CFT correspondence. 
Some open questions in 5d dS/4d CFT correspondence are briefly discussed.

\vfill

\ 

\newpage

\section{Introduction}

It became clear recently that holography should play the fundamental 
role in physics of 21th century. Moreover, there are significant changes 
in what we mean as holography by now. Indeed, roughly speaking 
we  expect that sometimes (at some, not strictly specified conditions) 
the classical volume physics may describe the quantum boundary physics
in less dimensions 
and vice-versa. The beatiful example of such holography is so-called 
AdS/CFT correspondence which occurs in string theory.
Considering the four-dimensional QFT, it says that five-dimensional AdS 
gravity is equivalent to some four-dimensional super Yang-Mills theory 
living on the boundary of AdS space which is gravitational ground state.

The connection between braneworld and AdS/CFT correspondence maybe
developed in
Brane New World construction
\cite{HHR,NOZ}. 
The quantum creation of  braneworld thanks to  conformal 
anomaly of four-dimensional quantum 
fields living on the brane occurs in such scenario.  

Due to general character of holographic principle it is expected that 
it should be realized in different ways. In particular, one can change the
bulk
(AdS) space and try to realize holography for another manifolds.
One of simplest candidates for bulk (apart from AdS)  is constant curvature
deSitter space. Then, holographic principle predicts that it should occur 
dS/CFT correspondence
\cite{strominger, desitter}.

The reason why AdS/CFT can be expected is the isometry of 
$d+1$-dimensional anti-de Sitter space, which is $SO(d,2)$ 
symmetry. It is identical with the conformal symmetry of 
$d$-dimensional Minkowski space. We should note, however,  
$d+1$-dimensional de Sitter space has the isometry of 
$SO(d+1,1)$ symmetry, which can be a conformal symmetry of 
$d$-dimensional Euclidean space. Then it might be natural to 
expect the correspondence between $d+1$-dimensional de Sitter 
space and $d$-dimensional euclidean conformal symmetry. 
In fact, the metric of 
$D=d+1$-dimensional anti de Sitter space (AdS) is given by
\be
\label{AdSm}
ds_{\rm AdS}^2=dr^2 + \e^{2r}\left(-dt^2 + \sum_{i=1}^{d-1}
\left(dx^i\right)^2\right)\ .
\ee
In the above expression, the boundary of AdS lies at 
$r=\infty$. If one exchanges the radial coordinate $r$ 
and the time coordinate $t$, we obtain the metric of the 
de Sitter space (dS): 
\be
\label{dSm}
ds_{\rm dS}^2=-dt^2 + \e^{2t}\sum_{i=1}^d
\left(dx^i\right)^2\ .
\ee
Here $x^d=r$. Then there is a boundary at $t=\infty$, where the 
Euclidean conformal field theory (CFT) can live and one expects 
dS/CFT correspondence as one more manifestation of holographic principle.
This may be very important as there are indications that our Universe has 
de Sitter phase in the past and in the future. Then, there appears very 
nice way to formulate some de Sitter gravitational physics in terms of 
the boundary QFT physics and vice-versa.

In the present contribution based mainly on ref.\cite{inf} we  
 consider the possibility of classical and  quantum creation of 
the inflationary brane in de Sitter bulk space in frames of mechanism of
refs. \cite{HHR,NOZ}.  
Moreover, the content of quantum fields on the brane may be chosen in such
a way, that it corresponds to euclidean CFT dual to 5d dS bulk space.
In this sense, one can understand that quantum creation
of dS brane-world occurs in frames of dS/CFT correspondence.
This is quite consistent, as is known (even in the absence of explicit CFT
dual to 5d dS space) that holographic conformal anomaly from 5d dS space is
proportional to the one of 4d super Yang-Mills theory\cite{dscft}. 
It is interesting that brane de Sitter gravity  (despite the fact 
that bulk represents not AdS but dS space) may be localized 
due to proposed dS/CFT correspondence.

\section{Classical and quantum de Sitter braneworlds}

The metric of  5 dimensional Euclidean de Sitter space 
that is 5d sphere is given by
\be
\label{dSi}
ds^2_{{\rm S}_5}=dy^2 + l^2 \sin^2 {y \over l}d\Omega^2_4\ .
\ee
Here $d\Omega^2_4$ describes the metric of ${\rm S}_4$ 
with unit radius. The coordinate $y$ is defined in 
$0\leq y \leq l\pi$. 
One also assumes the brane 
lies at $y=y_0$ 
and the bulk space is given by gluing two regions 
given by $0\leq y < y_0$. 

We start with the action $S$ which is the sum of 
the Einstein-Hilbert action $\SEH$ with positive 
cosmological constant, the Gibbons-Hawking 
surface term $\SGH$,  the surface counter term $S_1$\footnote{
The coefficient of $S_1$ cannot be determined from the condition of 
finiteness of the action as in AdS/CFT. However, using the 
renormailzation group method as in \cite{BVV}  this 
coefficient can be determined uniquely \cite{dscft}. } 
and the trace anomaly induced action 
${\cal W}$: 
\bea
\label{Stotal}
&& S=\SEH + \SGH + 2 S_1 + {\cal W}\ ,\quad 
\SEH={1 \over 16\pi G}\int d^5 x \sqrt{\gfv}\left(R_{(5)} 
 - {12 \over l^2}\right)\ , \nn
&& \SGH={1 \over 8\pi G}\int d^4 x \sqrt{\gfr}\nabla_\mu n^\mu 
\ ,\quad S_1= -{3 \over 8\pi Gl}\int d^4 x \sqrt{\gfr} \ ,\nn
&&{\cal W}= b \int d^4x \sqrt{\widetilde g}\widetilde F A 
 + b' \int d^4x \sqrt{\widetilde g}
\left\{A \left[2 {\wlBox}^2 
+\widetilde R_{\mu\nu}\widetilde\nabla_\mu\widetilde\nabla_\nu 
 - {4 \over 3}\widetilde R \wlBox^2 \right.\right. \nn
&& \left.\left. \qquad 
+ {2 \over 3}(\widetilde\nabla^\mu \widetilde R)\widetilde\nabla_\mu
\right]A 
+ \left(\widetilde G - {2 \over 3}\wlBox \widetilde R
\right)A \right\} \nn
&& \qquad -{1 \over 12}\left\{b''+ {2 \over 3}(b + b')\right\}
\int d^4x \sqrt{\widetilde g} \left[ \widetilde R - 6\wlBox A 
 - 6 (\widetilde\nabla_\mu A)(\widetilde \nabla^\mu A)
\right]^2 \ .
\eea 
Here the quantities in the  5 dimensional bulk spacetime are 
specified by the suffices $_{(5)}$ and those in the boundary 4 
dimensional spacetime  by $_{(4)}$. 
The factor $2$ in front of $S_1$ in (\ref{Stotal}) is coming from 
that we have two bulk regions which 
are connected with each other by the brane. 
In (\ref{Stotal}), $n^\mu$ is 
the unit vector normal to the boundary. In (\ref{Stotal}), 
one chooses the 4 dimensional boundary metric as 
\be
\label{tildeg}
\gfr_{\mu\nu}=\e^{2A}\tilde g_{\mu\nu}
\ee 
and we specify the 
quantities with $\tilde g_{\mu\nu}$ by using $\tilde{\ }$. 
$G$ ($\tilde G$) and $F$ ($\tilde F$) are the Gauss-Bonnet
invariant and the square of the Weyl tensor.

In the effective action (\ref{actions2}) induced by brane quantum 
matter, in general, with $N$ real scalar, $N_{1/2}$ 
Dirac spinor, $N_1$ vector 
fields, $N_2$  ($=0$ or $1$) gravitons and $N_{\rm HD}$ higher 
derivative conformal scalars, $b$, $b'$ and $b''$ are
\bea
\label{bs}
&& b={N +6N_{1/2}+12N_1 + 611 N_2 - 8N_{\rm HD} 
\over 120(4\pi)^2}\ ,\nn 
&& b'=-{N+11N_{1/2}+62N_1 + 1411 N_2 -28 N_{\rm HD} 
\over 360(4\pi)^2}\ ,\quad b''=0\ .
\eea

For typical examples motivated by AdS/CFT (and presumbly by dS/CFT 
because central charges are the same in AdS/CFT or dS/CFT)
correspondence 
one has:
a) ${\cal N}=4$ $SU(N)$ SYM theory 
$b=-b'={N^2 -1 \over 4(4\pi )^2}$, 
b) ${\cal N}=2$ $Sp(N)$ theory 
$b={12 N^2 + 18 N -2 \over 24(4\pi)^2}$, 
$b'=-{12 N^2 + 12 N -1 \over 24(4\pi)^2}$. 
 Note that $b'$ is negative in the above cases.

The metric of ${\rm S}_4$ with the unit radius is given by
\be
\label{S4metric1}
d\Omega^2_4= d \chi^2 + \sin^2 \chi d\Omega^2_3\ .
\ee
Here $d\Omega^2_3$ is described by the metric of 3 dimensional 
unit sphere. If one changes the coordinate $\chi$ to 
$\sigma$ by $\sin\chi = \pm {1 \over \cosh \sigma}$, 
one obtains\footnote{
If we Wick-rotate the metric by $\sigma\rightarrow it$, we 
obtain the metric of de Sitter space:
\[
d\Omega^2_4\rightarrow ds_{\rm dS}^2
= {1 \over \cos^2 t}\left(-dt^2 + d\Omega^2_3\right)\ .
\]
}
\be
\label{S4metric2}
d\Omega^2_4= {1 \over \cosh^2 \sigma}\left(d \sigma^2 
+ d\Omega^2_3\right)\ .
\ee
Then one assumes 
the metric of 5 dimensional space time as follows:
\be
\label{metric1}
ds^2=dy^2 + \e^{2A(y,\sigma)}\tilde g_{\mu\nu}dx^\mu dx^\nu\ ,
\quad \tilde g_{\mu\nu}dx^\mu dx^\nu\equiv l^2\left(d \sigma^2 
+ d\Omega^2_3\right)
\ee
and one identifies $A$ and $\tilde g$ in (\ref{metric1}) with those in 
(\ref{tildeg}). Then $\tilde F=\tilde G=0$, 
$\tilde R={6 \over l^2}$ etc. 
Due to the assumption (\ref{metric1}), the actions in (\ref{Stotal}) 
have the following forms:
\bea
\label{actions2}
&& \SEH= {l^4 V_3 \over 16\pi G}\int dy d\sigma \left\{\left( -8 
\partial_y^2 A - 20 (\partial_y A)^2\right)\e^{4A} \right. \nn
&& \qquad \left. +\left(-6\partial_\sigma^2 A 
 - 6 (\partial_\sigma A)^2 
+ 6 \right)\e^{2A} - {12 \over l^2} \e^{4A}\right\} \nn
&& \SGH= {l^4 V_3 \over 2\pi G}\int d\sigma \e^{4A} 
\partial_y A \ ,\quad 
S_1= - {3l^3 V_3 \over 8\pi G}\int d\sigma \e^{4A} \nn
&& {\cal W}= V_3 \int d\sigma \left[b'A\left(2\partial_\sigma^4 A
 - 8 \partial_\sigma^2 A \right) 
 - 2(b + b')\left(1 - \partial_\sigma^2 A 
 - (\partial_\sigma A)^2 \right)^2 \right]\ .
\eea
Here $V_3$ is the volume or area of the unit 3 sphere. 

In the bulk, one obtains the following equation of motion 
from $\SEH$ by the variation over $A$:
\be
\label{eq1}
0= \left(-24 \partial_y^2 A - 48 (\partial_y A)^2 
 - {48 \over l^2}
\right)\e^{4A} + {1 \over l^2}\left(-12 \partial_\sigma^2 A 
- 12 (\partial_\sigma A)^2 + 12\right)\e^{2A}\ ,
\ee
which corresponds to one of the Einstein equations. 
Then one finds solutions, $A_S$, which correspond to 
the metric  dS$_5$ in (\ref{dSi}) with (\ref{S4metric2}). 
\be
\label{blksl}
A=A_S=\ln\sin{y \over l} - \ln \cosh\sigma\ .
\ee

On the brane at the boundary, 
one gets the following equation:
\bea
\label{eq2}
0&=&{48 l^4 \over 16\pi G}\left(\partial_y A - {1 \over l}
\right)\e^{4A}
+b'\left(4\partial_\sigma^4 A - 16 \partial_\sigma^2 A\right) \nn
&& - 4(b+b')\left(\partial_\sigma^4 A + 2 \partial_\sigma^2 A 
 - 6 (\partial_\sigma A)^2\partial_\sigma^2 A \right)\ .
\eea
We should note that the contributions from $\SEH$ and $\SGH$ are 
twice from the naive values since we have two bulk regions which 
are connected with each other by the brane. 
Substituting the bulk solution $A=A_S$ in (\ref{blksl}) into 
(\ref{eq2}) and defining the radius $R$ of the brane by
$R\equiv l\sin{y_0 \over l}$, one obtains
\be
\label{slbr2}
0={1 \over \pi G}\left({1 \over R}\sqrt{1 - {R^2 \over l^2}}
 - {1 \over l}\right)R^4 + 8b'\ .
\ee
One sees that eq.(\ref{slbr2}) does not depend on $b$. 
First we should note $0\leq R\leq l$ by  definition. 
Even in the case that there is no quantum contribution 
from the matter on the brane, that is, $b'=0$, Eq.(\ref{slbr2}) 
has a solution:
\be
\label{Csol}
R^2=R_0^2\equiv 
{l^2 \over 2}\ \mbox{or}\ {y_0 \over l}={\pi \over 4}, 
{3\pi \over 4}\ .
\ee
In Eq.(\ref{slbr2}), the first term 
${R^3 \over \pi G}\sqrt{1 - {R^2 \over l^2}}$ 
corresponds to the gravity, which makes the radius $R$ larger. 
On the other hand, the second term 
$-{R^4 \over \pi Gl}$ corresponds to the tension, which makes 
$R$ smaller. When $R<R_0$,  gravity becomes larger than the 
tension and when $R>R_0$, vice versa. Then both of the solutions in 
(\ref{Csol}) are stable. Although it is not clear from 
(\ref{slbr2}), $R=l$ (${y \over l}={\pi \over 2}$) corresponds 
to the local maximum. 
Hence, the possibility of creation of inflationary brane in 
de Sitter bulk 
is possible already on classical level, even in situation when brane 
tension is fixed by holographic RG. That is qualitatively different from 
the case of AdS bulk where only quantum effects led to creation of
inflationary 
brane \cite{NOZ,HHR} (when brane tension was not free 
parameter). Of course, if brane tension is taken to be arbitrary, i.e. the
coefficient of  $S_1$ is not fixed then there appears more possibilities.

Let us make several remarks about properties of dS brane-world.
There is an excellent explanation \cite{HHR}  why gravity 
is trapped on the brane in the AdS spacetime. This uses AdS$_5$/CFT$_4$ 
correspondence. This can be generalized to 
the brane in dS spacetime by using proposed dS/CFT correspondence. 

In \cite{BBM} it has been shown that  the bulk action diverges
in de Sitter space when we substitute the classical 
solution, which is the fluctuation around the de Sitter space 
in (\ref{dSm}). In other words,  counterterms are necessary again. The 
divergence 
occurs since the volume of the space diverges when $t\rightarrow 
\infty$ (or $t\rightarrow -\infty$ after replacing $t$ by $-t$ 
in another patch). Then we should put the counterterms on the 
space-like branes which lie at $t\rightarrow \pm\infty$. 
Therefore dS/CFT correspondence should be given by
\bea
\label{lc1}
&& \e^{-W_{\rm CFT}}=\int [dg][d\varphi]\e^{-S_{\rm dS\,grav}}\ , 
\quad S_{\rm dS\,grav}=\SEH + \SGH + S_1 + S_2 + \cdots, \nn 
&& \SEH={1 \over 16\pi G}\int d^5 x \sqrt{-\gfv}\left(R_{(5)} 
 - {12 \over l^2} + \cdots \right)\ , \nn 
&& \SGH={1 \over 8\pi G}\int_{M_4^+ + M_4^-} d^4 x 
\sqrt{\gfr}\nabla_\mu n^\mu, \\
&& S_1= {3 \over 8\pi G l}\int_{M_4^+ + M_4^-} d^4 x 
\sqrt{\gfr}\ , \quad 
S_2= {l \over 32\pi G }\int_{M_4^+ + M_4^-} d^4 x 
\sqrt{\gfr}\left(R_{(4)} + \cdots \right)\ ,  \cdots \ . 
\nonumber
\eea
Here $S_1$, $S_2$, $\cdots$ correspond to the surface counterterms, 
which cancell the divergences in the bulk action and $M_4^\pm$ 
expresses the boundary at $t\rightarrow \pm \infty$.  

Let us consider two copies of the de Sitter spaces dS$_{(1)}$ 
and dS$_{(2)}$. We also put one or two of the space-like 
branes, which can be 
regarded as boundaries  connecting the two bulk de Sitter 
spaces, at finite $t$. Then if one takes the 
following action $S$ instead of $S_{\rm dS\,grav}$, 
\be
\label{lc2}
S=\SEH + \SGH + 2S_1=S_{\rm dS\,grav} + S_1 - S_2 - \cdots,
\ee
we obtain the following boundary theory in terms of 
the partition function:
\bea
\label{lc3}
&& \int_{{\rm dS_5^{(1)} + dS_5^{(1)}} +M_4^+ + M_4^-} 
[dg][d\varphi]\e^{-S} 
= \left(\int_{{\rm dS}_5} [dg][d\varphi]\e^{-\SEH - \SGH - S_1} 
\right)^2 \nn
&& \quad =\e^{2S_2 + \cdots}\left(\int_{{\rm dS}_5} [dg][d\varphi]
\e^{-S_{\rm grav}} \right)^2 =\e^{-2W_{\rm CFT}+2S_2 + \cdots}\ .
\eea
Since $S_2$ can be regarded as the Einstein-Hilbert action on 
the boundary, the gravity on the boundary becomes dynamical. 
In other words, there is strong indication that  braneworld model under
consideration at 
some conditions may be effectively described by 4d gravity interacting with 
some gauge theory. In fact, the explicit proof that indeed gravity trapping 
occurs on dS brane in 5d dS bulk is given in \cite{brevik} and standard
newton potential is induced there \cite{newton}. This is extremely 
powerful argument in favor of dS/CFT correspondence.

Now we consider the quantum effects ($b'\neq 0$ case) on 
the brane in (\ref{slbr2}). Let us define a function 
$F(R^2)$ as follows:
\be
\label{F}
F(R^2)={1 \over \pi G}\left({1 \over R}\sqrt{1 - {R^2 \over l^2}}
 - {1 \over l}\right)R^4 \ .
\ee
Then one can easily find
\bea
\label{prF}
&& F(0)=F\left({l^2 \over 2}\right)=0 \ ,\quad 
F(l^2)=-{l^3 \over \pi G} \ ,\nn
&& F(R^2)\begin{array}{ll}
 >0 \quad & \mbox{when}\ 0<R^2<{l^2 \over 2} \\
 <0 \quad & \mbox{when}\ {l^2 \over 2}<R^2 \leq l^2 \\
\end{array} \ .
\eea
The function $F(R^2)$ has a maximum 
\be
\label{mF0}
F=F_m\equiv {l^3 \over 16\pi G}\left(-26 
+ 35 \sqrt{1-{9 \over 50}}\right)
\ee
when 
\be
\label{mF}
R^2=R_m^2\equiv {5l^2 \over 4}\left(1-\sqrt{1-{9 \over 50}}
\right)<{l^2 \over 2}\ .
\ee
The above results tell 
\begin{enumerate}
\item When $-8b'> F_m$ or $-8b' < -{l^3 \over \pi G}$, there 
is no solution in Eq.(\ref{slbr2}). That is, the quantum effect 
exhibits the creation of the inflationary brane world. 
\item When $0<-8b'< F_m$, there appear two solutions in 
(\ref{slbr2}). The solution with larger radius $R$ corresponds to 
the classical one in (\ref{Csol}) but the radius $R$ in the 
solution is smaller then that in the classical one. 
In other words, quantum effects try to prevent inflation. The solution 
with smaller radius can be regarded as the solution  generated 
by only quantum effects on the brane. Anyway the radii $R$ in 
both of the solutions are smaller than that in the classical one 
(\ref{Csol}). Since ${1 \over R}$ corresponds to the rate of the 
expansion of the universe when S$_4$ is Wick-rotated into 4d de 
Sitter space, the quantum effect makes the rate larger. 
\item When $0>-8b'>-{l^3 \over \pi G}$, which is rather exotic 
case since usualy $b'$ is negative as in (\ref{bs}), 
Eq.(\ref{slbr2}) has unique solution corresponding to the 
solution in the classical case (\ref{Csol}) and the quantum 
effect on the brane makes the radius $R$ larger. 
\end{enumerate}

The de Sitter brane may be thought as inflationary brane.
 Moreover, we got the alternative 
description of the dS braneworld as some 4d gravity 
with matter. 
Then principal possibility of the end of brane inflation maybe established
\cite{inf}.

\section{Discussion}

The consideration of 5d deSitter braneworld above strongly 
indicates to 5d dS/4d CFT correspondence. In many respects 
as is shown in refs.\cite{5dds} it is similar to 5d AdS/4d CFT correspondence.
Moreover, similar questions as in AdS/CFT correspondence maybe addressed
also in dS/CFT correspondence.

Unfortunately, unlike to string-motivated AdS/CFT duality 
one cannot present yet the consistent and satisfactory model of CFT 
(for which even central charge is known \cite{dscft}) dual to 5d dS space.
In many respects this slow down the development of dS/CFT 
correspondence. It is quite possible also that such situation indicates 
to necessity of essential modification of what we mean as CFT now 
(see one explicit example in \cite{desitter}). 
In particular, it is expected that such dual CFT may have some
 problems with unitarity, and (or) with tachyons. From another side,
many properties of such CFT are known from gravitational description.
In particulary, even Wilson loop \cite{wilson} in such CFT maybe found 
explicitly. Unlike to AdS/CFT the quark-antiquark potential indicates to
the possibility of confinement. In this respect, dS/CFT correspondence 
may favor QCD better that AdS holographic description.

\section{Acknoweledgements} I am very grateful to S. Nojiri with whom
most of results discussed in this talk are obtained.

\end{document}